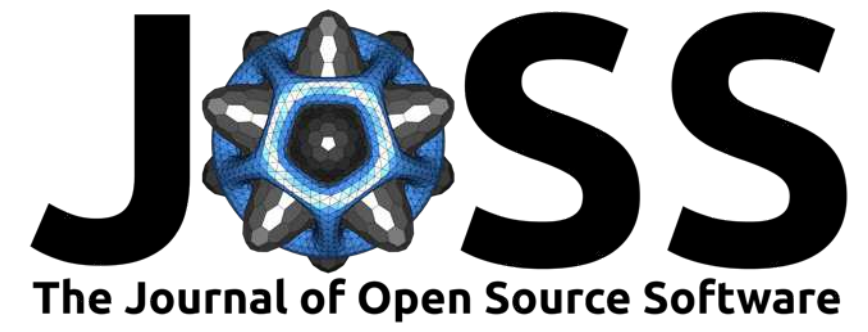

# OGRe: An Object-Oriented General Relativity Package for Mathematica

**Barak Shoshany**[1]

**1** Department of Physics, Brock University, 1812 Sir Isaac Brock Way, St. Catharines, Ontario, L2S 3A1, Canada

## Summary

OGRe is a modern Mathematica package for differential geometry and tensor calculus. It can be used in a variety of contexts where tensor calculations are needed, in both mathematics and physics, but it is especially suitable for general relativity — the field of physics where tensors are most commonly and ubiquitously used. Whether the user is doing cutting-edge research in general relativity or just making first steps in learning the theory, the ability to manipulate tensors and perform tensor calculations quickly, easily, and intuitively will greatly simplify and accelerate their work.

Tensors are abstract geometrical structures, which describe curved spaces and objects within these spaces. In principle, it is possible to perform calculations with the abstract tensors themselves, and this is often done in pure mathematics. However, in practice, one usually represents a tensor as a set of individual components — similarly to how an abstract vector is just an arrow, but concrete calculations usually involve representing the vector as a list of components. The mathematical details are given in the statement of need below.

Unfortunately, tensor calculations are notoriously complicated and prone to errors. Tensors have many individual components, and operations on tensors involve manipulating and combining the components of one or more tensors in convoluted ways. Furthermore, combining several tensors requires the representations of each of the tensors involved to be compatible with each other according to strict rules.

OGRe is designed to simplify the complexities of tensor calculations. This is done using an object-oriented programming approach, taking advantage of principles such as encapsulation and class invariants to eliminate the possibility of user error. A single tensor object in OGRe contains the components of the tensor in different representations, as well as metadata such as the type of the tensor and the symbol used to represent it in equations.

To construct a new object, the user only needs to enter the tensor's components — a multidimensional array of numbers, symbols, and/or functions — in one representation. Other representations will then be calculated automatically by OGRe as needed, by transforming the initial components behind the scenes using the appropriate rules.

Operations on tensors are performed by the user abstractly, without specifying which representations to use. OGRe's algorithm will automatically determine and use the correct combination of tensor representations needed for the specific operation, no matter how complicated the operation is. This ensures that the user cannot mistakenly perform "illegal" operations, that is, combine tensors of non-compatible representations.

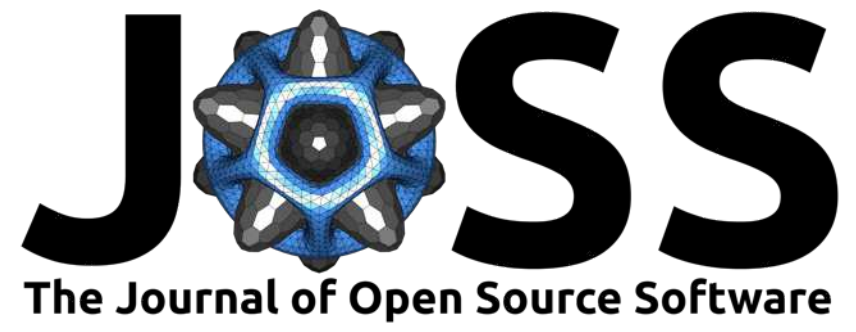

## Statement of need

Tensors are defined in a coordinate-independent way as multi-linear maps on vectors and covectors — where covectors are linear maps from vectors to the real numbers. A tensor which acts on $p$ covectors and $q$ vectors is said to be of rank $(p,q)$. Given a choice of coordinate system, a tensor can be represented as a multi-dimensional array. The components of this array can be described using a set of $p+q$ indices, with $p$ upper indices and $q$ lower indices, e.g. $T^{\mu_1\cdots\mu_p}{}_{\nu_1\ldots\nu_q}$ — where each of the indices $\{\mu_1\ldots\mu_p,\nu_1\ldots\nu_q\}$ takes values from 1 to the number of dimensions in the space. A rank $(0,0)$ tensor is a scalar, a rank $(1,0)$ tensor is a vector, and a rank $(0,1)$ tensor is a covector.

The most important use of tensors is in the context of curved spaces, notably in general relativity, where gravity is described using a curved 4-dimensional spacetime. The curvature is encoded in a special rank $(0,2)$ tensor called the metric. The metric can be used to raise and lower indices, that is, turn a lower index into an upper index or vice-versa. This means that for each non-negative integer $k$, all the spaces of rank $(p,q)$ tensors with $p+q=k$ are isomorphic. Therefore, we can define a more general notion of abstract tensors of rank $k$, whose representations have $k$ indices in total, but with a different number of upper vs. lower indices for each representation. One rank $k$ tensor will thus have many different representations, depending both on the coordinate system and the index configuration.

Transforming a tensor representation from one coordinate system to another is done by taking complicated combinations of the tensor's components with the Jacobian of the coordinate transformation. Transforming from one index configuration to another is done similarly, by taking complicated combinations of the components with the metric. Given that tensor representations typically have dozens or even hundreds of individual components, this can be a very complicated task.

Operations on one or more tensors can be even more complicated, since the representations of the different tensors have to match. For example, addition of tensors may only be performed component-by-component if all tensors are in the exact same representation. On the other hand, contraction of an index of one tensor with an index of another tensor, which is a generalization of the notion of inner product, requires choosing the representations of the tensors such that one index being contracted is upper and the other is lower.

When doing such calculations by hand, it is quite easy to lose track and make mistakes — as every student of differentia geometry and general relativity inevitably discovers. Computer algebra systems, such as Mathematica (Wolfram Research, Inc., 2021), are thus indispensable for doing tensor calculations (MacCallum, 2018). They save considerable time and effort that would have been spent performing the calculations by hand, but more importantly, they ensure that the final results are free of errors.

However, as Mathematica cannot perform non-trivial tensor calculations out of the box, one has to define each operation individually with the correct combination of components in the correct representations, which is by itself a difficult and delicate task. Therefore, various Mathematica packages, most notably xAct (Martín-García, 2021), have been created to provide a higher-level implementation of tensors. These packages are very powerful, and are an indispensible tool for many researchers, but they also tend to have complex and unintuitive interfaces, which can be overwhelming to new users.

OGRe is intended to be intuitive, user-friendly, and easy to learn and use, while also being robust and rich in features. It is designed with elegance and simplicity in mind, and comes with built-in tools for displaying tensors and their components in instructive and visually pleasing ways. Furthermore, unlike other packages, OGRe was written from scratch in Mathematica 12, and makes ample use of many new Mathematica features for increased performance, functionality, and ease of use.

The package is designed to be accessible even to users who do not have much experience with Mathematica and/or general relativity, while also being robust and rich in features. As a result, it is equally suitable for both experienced and novice researchers.

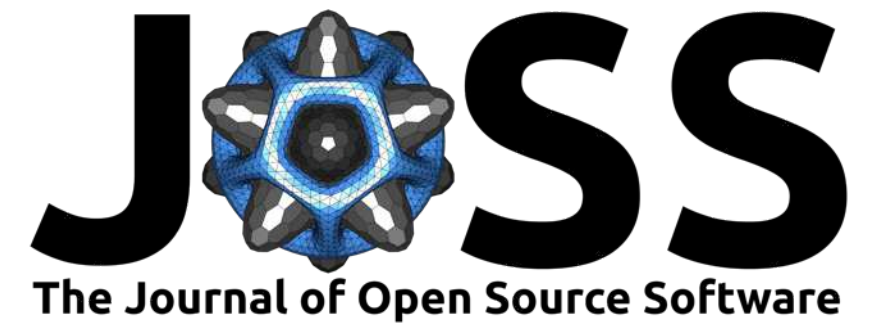

The reader is invited to read more about tensors in differential geometry in Kobayashi & Nomizu (1996a) and Kobayashi & Nomizu (1996b), their use in physics in Baez & Muniain (1994), Frankel (2011), and Nakahara (2018), and their use in general relativity in Carroll (2019) and Wald (2010).

## Overview of features

- Define coordinate systems and the transformation rules between them. Tensor components are then transformed automatically between coordinates behind the scenes as needed.
- Each tensor is associated with a specific metric. Tensor components are then transformed automatically between different index configurations, raising and lowering indices behind the scenes as needed.
- Display any tensor in any index configuration and coordinate system, either in vector/matrix form or as a list of all unique non-zero elements. Metrics can also be displayed as a line element.
- Automatically simplify tensor components, optionally with user-defined simplification assumptions. Simplifications can be parallelized for a significant performance boost.
- Export tensors to a Mathematica notebook or to a file, so they can later be imported into another Mathematica session without having to redefine them from scratch.
- Highly customizable. User settings are exported and imported along with the tensors. Some settings are persistent between sessions.
- Easily calculate arbitrary tensor formulas using any combination of addition, multiplication by scalar, trace, contraction, partial derivative, and covariant derivative.
- Built-in modules for calculating the Christoffel symbols (Levi-Civita connection), Riemann tensor, Ricci tensor and scalar, Einstein tensor, curve Lagrangian, and volume element from a metric.
- Calculate the geodesic equation, in two different ways: from the Christoffel symbols or from the curve Lagrangian.
- Built with speed and performance in mind, using optimized algorithms designed specifically for this package.
- Fully portable. Can be imported directly from the web into any Mathematica notebook, without downloading or installing anything. Integrates seamlessly with the Wolfram Cloud.
- Clear and detailed documentation, with many examples, in both Mathematica notebook and PDF format. Detailed usage messages are also provided.
- Open source. The code is extensively documented; please feel free to fork and modify it as you see fit.
- Under continuous and active development. Bug reports and feature requests are welcome, and should be made via GitHub issues.

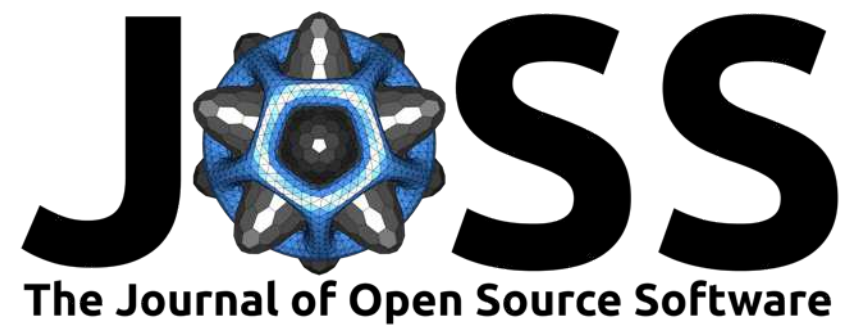